\documentclass[aps,pra,reprint,amsmath,amssymb,superscriptaddress,nobibnotes, longbibliography]{revtex4-1}
\usepackage{graphicx,SIunits}
\usepackage{bm}     
\usepackage{hyperref} 
\usepackage{dsfont}    
\usepackage{color}

\usepackage{amsmath}
\usepackage{amsthm}
\usepackage{amssymb}
\usepackage{tikz}

\usetikzlibrary{decorations.pathreplacing}
\usepackage{braket}
\usepackage{dsfont}

\usepackage{color}

\begin{document}
	
	\title{Optimal multiple-phase estimation with multi-mode NOON states against photon loss}
	\author{Min Namkung}
	\affiliation{Center for Quantum Information, Korea Institute of Science and Technology (KIST), Seoul, 02792, Korea}
	
	\author{Dong-Hyun Kim}
	\affiliation{Center for Quantum Information, Korea Institute of Science and Technology (KIST), Seoul, 02792, Korea}
	\affiliation{Department of Physics, Yonsei University, Seoul 03722, Korea}
	
	\author{Seongjin Hong}
	\affiliation{Department of Physics, Chung-Ang University, Seoul 06974, Korea}
	
	\author{Yong-Su Kim}
	\affiliation{Center for Quantum Information, Korea Institute of Science and Technology (KIST), Seoul, 02792, Korea}
	\affiliation{Division of Nanoscience and Technology, KIST School, Korea University of Science and Technology, Seoul 02792, Korea}

        \author{Changhyoup Lee}
        \email{changhyoup.lee@gmail.com}
        \affiliation{Korea Research Institute of Standards and Science, Daejeon 34113, Korea}
	
	\author{Hyang-Tag Lim}
	\email{hyangtag.lim@kist.re.kr}
	\affiliation{Center for Quantum Information, Korea Institute of Science and Technology (KIST), Seoul, 02792, Korea}
	\affiliation{Division of Nanoscience and Technology, KIST School, Korea University of Science and Technology, Seoul 02792, Korea}

\date{\today} 
	
\begin{abstract}
Multi-mode NOON states can quantum-enhance multiple-phase estimation in the absence of photon loss. However, a multi-mode NOON state is known to be vulnerable to photon loss, and its quantum-enhancement can be dissipated by lossy environment. In this work, we demonstrate that a quantum advantage in estimate precision can still be achieved in the presence of photon loss. This is accomplished by optimizing the weights of the multi-mode NOON states according to photon loss rates in the multiple modes, including the reference mode which defines the other phases. For practical relevance, we also show that photon-number counting via a multi-mode beam-splitter achieves the useful, albeit sub-optimal, quantum advantage. We expect this work to provide valuable guidance for developing quantum-enhanced multiple-phase estimation techniques in lossy environments.
\end{abstract}

\maketitle

\section{Introduction}
Quantum resources can enhance performance of precisely estimating multiple parameters over classical counterparts. For this reason, most studies have focused on finding optimal quantum resources achieving the ultimate quantum limit in multiple-parameter estimation \cite{e.polino,m.szczy,j.f.haase}, including the use of the Greenberger-Horne-Zeilinger states \cite{l.z.liu,s.-r.zhao}, single-photon Fock states \cite{e.polino2}, squeezed states \cite{y.xia,r.schnabel,b.j.lawrie,x.guo,m.gessner,c.oh,s.-i.park}, entangled coherent states \cite{j.joo,j.liu,s.-y.lee}, Holland-Burnett states \cite{z.su,m.a.ciampini,g.y.xiang}, and multi-mode NOON states \cite{p.c.humphreys,l.zhang,l.zhang2,s.hong,s.hong2,j.urrehman}. Among them, the multi-mode NOON states are particularly known to achieve the enhanced estimate precision outperforming the other probe states in the mean photon number for multiple-phase estimation, consequently beating the standard quantum limit (SQL) \cite{p.c.humphreys}. From this fact, it is conceivable that multi-mode NOON states could serve as a useful resource for developing novel quantum sensing technology. 

However, when a multi-mode NOON state is exposed to noisy or lossy environment, it does not always provide the quantum advantage in precisely estimating multiple-phases \cite{r.demkowicz,u.doner,r.demkowicz2,j.kolodynski,m.kacprowicz}. Disadvantage caused by photon loss when using multi-mode NOON states in multiple-phase estimation becomes serious  as the photon number $N$ increases \cite{u.doner}. Fortunately, the number of modes in a multi-mode NOON state is known to be a quantified resource for improving the multiple-phase estimation in lossless case \cite{j.-d.yue}. Thus, it is utmost importance to verify whether a quantum advantage is also improved by increasing the number of modes in a multi-mode NOON state against photon loss, even in the case that a reference mode is more severely exposed to lossy  environments than the other modes.

In this work, we propose an optimal multiple-phase estimation scheme using the weighted multi-mode NOON states in the presence of photon loss. Instead of actively correcting errors, which requires more resources, we rather optimize the probe states to minimize the impact of photon loss in our scheme. Analytically deriving the estimation precision for the proposed scheme, we investigate  the robustness of the scheme against photon loss and compare it with the scheme using a particular type of the multi-mode NOON state proposed in Ref.~\cite{p.c.humphreys}, which is known to be optimal in the lossless case. A quantum advantage is also studied in comparison with the SQL set by a scheme using the weighted multi-mode pure coherent states. As a result, we find that our scheme is not only more robust against photon loss than the scheme in Ref.~\cite{p.c.humphreys}, but also exhibits a quantum advantage over the SQL even in the presence of loss. These findings are elaborated in more details for the cases of three- and four-mode two-photon NOON states that have been experimentally demonstrated \cite{s.hong,s.hong2}, followed by the generalization of the scheme with increasing the numbers of modes and photons. The proposed schemes are also investigated with a measurement scheme with general structure consisting of a multi-mode beam-splitter and multiple photon-number-resolving detectors (PNRDs), offering a practical means to achieve the quantum advantage in multiple-phase estimation under a lossy environment. We expect that our results and methodology can be used for various studies in robust quantum metrology, given that loss and decoherence are inevitable. 

\section{Multiple-phase estimation scheme under lossy environment}

\begin{figure}[t]
\rightline{\includegraphics[width=8.5cm]{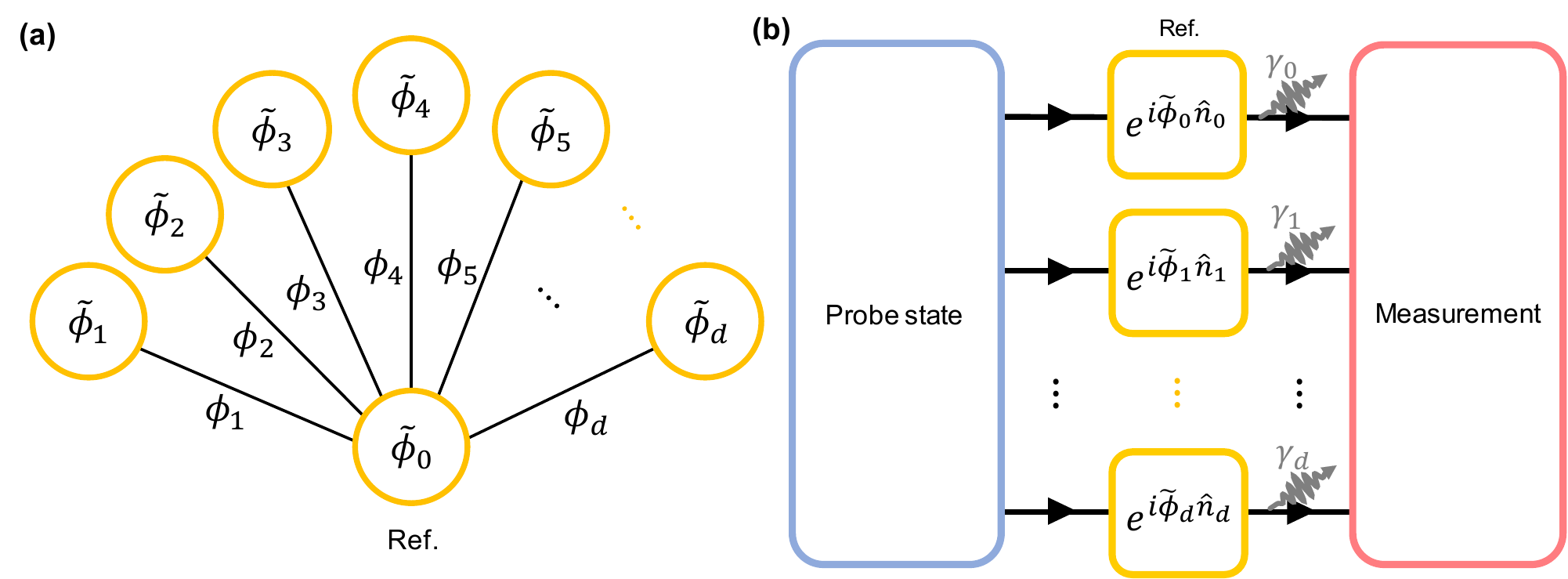}}
\caption{(a) Illustrative representation of the relative phases $\phi_j=\widetilde{\phi}_j-\widetilde{\phi}_0$ and (b) multiple-phase estimation using a $d+1$-mode probe state for estimating $d$ unknown relative phases $\{\phi_1,\cdots,\phi_d\}$ with a measurement. Without loss of generality, the mode 0 is considered as a reference. It is assumed that photon loss rate  in the mode $j$ is $\gamma_j\in[0,1]$.}
\centering
\label{fig1}
\end{figure}

\subsection{Scheme}
Let us consider a scheme for simultaneously estimating $d$ unknown relative phases $\boldsymbol{\phi}=\{\phi_1,\cdots,\phi_{d}\}$ with $\phi_j=\widetilde{\phi}_j-\widetilde{\phi}_0$ in the presence of photon loss as shown in Fig.~\ref{fig1}(a). There, we assume that the mode 0 is used as a reference upon which the phases of all the other modes are defined. We note that our estimation problem is equivalent to what was originally considered in Ref. \cite{p.c.humphreys}. For the multiple-phase estimation, we employ a ($d+1$)-mode NOON state probe written by
\begin{align}\label{NOON}
|\psi_{N}\rangle=\sum_{j=0}^{d}\sqrt{p_j}\frac{\hat{a}_j^{\dagger N}}{\sqrt{N!}}|0\rangle^{\otimes d+1},
\end{align}
 where $\hat{a}_j^\dagger$ is the photon creation operator in the mode $j$ and $p_j$ with $\sum_{j=0}^d p_j=1$ and $p_j\in[0,1]$ is the weight of the mode $j$. This probe state undergoes $d$ individual phase shifts described by a unitary operator
 $\hat{U}_{\boldsymbol{\widetilde{\phi}}}=\bigotimes_{j=0}^{d}\exp(i\widetilde{\phi}_j\hat{a}_j^\dagger\hat{a}_j)$ with a tuple $\boldsymbol{\widetilde{\phi}}=\{\widetilde{\phi}_0,\cdots,\widetilde{\phi}_d\}$ as illustrated in Fig.~\ref{fig1}(b). For practical relevance, we take into account photon loss that could occur in the transmission channels or inefficient detectors. The photon loss can be modeled by a fictitious beam-splitter represented by the mode transformation $\hat{a}_j^\dagger\rightarrow\sqrt{1-\gamma_j}\hat{a}_j^\dagger+\sqrt{\gamma_j}\hat{e}_j^\dagger$ \cite{p.kok} with photon loss rate $\gamma_j\in[0,1]$, where $\hat{e}_j^{\dagger}$ is a virtual vacuum input mode for describing environment. It can be shown, by the Kraus representation of photon loss, that photon losses occurring before and after the phase shifters are equivalent. This allows to consider only the photon loss taking place onto the phase-encoded probe state $|\psi_{N,\boldsymbol{\widetilde{\phi}}}\rangle=\hat{U}_{\boldsymbol{\widetilde{\phi}}}|\psi_N\rangle$ before measurement. The phase-encoded ($d+1$)-mode NOON state is thus transformed to the output state 
\begin{align}\label{lossy_NOON}
\hat{\rho}_{N,\boldsymbol{\widetilde{\phi}}}=\Lambda_{\gamma_0}\otimes\cdots\otimes\Lambda_{\gamma_{d}}(|\psi_{N,\boldsymbol{\widetilde{\phi}}}\rangle\langle\psi_{N,{\boldsymbol{\widetilde{\phi}}}}|),
\end{align}
where $\Lambda_{\gamma_j}$ is a photon loss channel in the mode $j$~\cite{m.grassl}. If $\gamma_j$ is zero, then the quantum channel $\Lambda_{\gamma_j}$ becomes an identity channel. Finally, the estimation of $d$ unknown relative phases is made based on the measurement results obtained by a measurement performed onto the output state of Eq.~(\ref{lossy_NOON}). 

\subsection{Estimation precision}
Denoting $|\Delta\phi_j|^2=\langle(\phi_{\mathrm{est},j}-\phi_j)^2\rangle$ as an estimation uncertainty of the unknown phase $\phi_j$, the total uncertainty of simultaneous estimation of $d$ phases can be defined as the sum of individual uncertainties, i.e., $\sum_{j=1}^{d}|\Delta\phi_j|^2$. When an unbiased estimator is used, the total uncertainty is lower-bounded by \cite{p.c.humphreys,j.urrehman,j.liu2}
\begin{align}\label{bound_sen}
    \sum_{j=1}^{d}|\Delta\phi_j|^2\ge\frac{\mathrm{Tr}\big[{F}_{\rm C}^{-1}(\boldsymbol{\phi})\big]}{\mu}\ge\frac{\mathrm{Tr}\big[{F}_{\rm Q}^{-1}(\boldsymbol{\phi})\big]}{\mu},
\end{align}
where $\mu$ is the number of measurement being repeated, $F_{\rm C}(\boldsymbol{\phi})$ is the Fisher information matrix (FIM), and $F_{\rm Q}(\boldsymbol{\phi})$ is the quantum Fisher information matrix (QFIM). Here, an FIM element is defined as 
\begin{align}\label{fim_def}
    {F}_{{\rm C}(j,k)}  =\sum_{\boldsymbol{l}}\frac{1}{P_{\boldsymbol{l}}}\left(\frac{\partial P_{\boldsymbol{l}}}{\partial\phi_j}\right)\left(\frac{\partial P_{\boldsymbol{l}}}{\partial\phi_k}\right),
\end{align}
with the conditional probability $P_{\boldsymbol{l}}$ to obtain a measurement outcome $\boldsymbol{l}$ conditioned on $\boldsymbol{\phi}$. QFIM is defined as \cite{j.liu2}
\begin{align}\label{qfim}
    {F}_{{\rm Q}(j,k)}=\frac{1}{2}\mathrm{Tr}\Big[\hat{\rho}_{N,\boldsymbol{\phi}}\left\{\hat{L}_j,\hat{L}_k\right\}\Big],
\end{align}
where $\{\cdot,\cdot\}$ denotes the anti-commutation and $\hat{L}_j$ is a symmetric logarithmic derivative operator with respect to an unknown phase $\phi_j$. In Eq.~(\ref{bound_sen}),  ${\rm Tr}\left[F_{\rm C}^{-1}(\boldsymbol{\phi})\right]$ and ${\rm Tr}\left[{F}_{\rm Q}^{-1}(\boldsymbol{\phi})\right]$ are called the Cramer-Rao bound (CRB) and the quantum Cramer-Rao bound (QCRB) (per a single measurement try), respectively. Note that the CRB is a lower bound of the total uncertainty determined for a given probe state and a chosen measurement scheme, whereas the QCRB is a further lower bound that would be obtained when an optimal measurement is employed for a given probe state. That is, the CRB becomes the QCRB if a chosen measurement scheme is optimal. We note that this attainability is possible in multiple-phase estimation. In general multiple-parameter estimation, there is a case that the CRB of an optimal measurement scheme does not achieve the QCRB \cite{j.liu2,l.pezze}. However, a measurement attaining the QCRB is experimentally challenging to implement in general. For example, when multiple phases are encoded in a multi-mode NOON state $|\psi_N\rangle$ in Eq.~(\ref{NOON}), this measurement is represented as a positive-operator-valued measure (POVM) including a projector $|\psi_N\rangle\langle\psi_N|$ \cite{p.c.humphreys,l.pezze}. The fact that the multi-mode NOON state is generated by nonlinear optics implies the difficulty in experimentally realizing the measurement \cite{s.hong,s.hong2,j.urrehman}. Thus, it is necessary to consider the CRB provided by a measurement experimentally configurable by linear optics. We also note that the CRB is asymptotically attained by using maximum likelihood estimator when $\mu\gg1$ \cite{l.pezze,c.w.helstrom}.

\section{Quantum advantage in terms of the QCRB}

\subsection{Minimum QCRB of multi-mode NOON states}
To evaluate the QCRB, we first derive the output state $\hat{\rho}_{N,\boldsymbol{\phi}}$ defined in Eq.~(\ref{lossy_NOON}) for an arbitrary photon number $N$ and phases $d$, as
\begin{align}\label{expl_NOON}
    \hat{\rho}_{N,\boldsymbol{\widetilde{\phi}}}=\left(\sum_{j=0}^{d}p_j(1-\gamma_j)^N\right)|{\psi}_{N,\boldsymbol{\widetilde{\phi}}}^{(\boldsymbol{\gamma})}\rangle\langle {\psi}_{N,\boldsymbol{\widetilde{\phi}}}^{(\boldsymbol{\gamma})}|+\hat{\sigma}_{N}.
\end{align}
Here, $\hat{\sigma}_{N}$ is a phase-independent non-negative density matrix (see Appendix A) diagonalized in the Fock state basis up to $(N-1)$, and $|{\psi}_{N,\boldsymbol{\widetilde{\phi}}}\rangle$ is a lossy multi-mode NOON state in the $N$-photon manifold written as
\begin{eqnarray}\label{NOON2}
    |{\psi}_{N,\boldsymbol{\widetilde{\phi}}}^{(\boldsymbol{\gamma})}\rangle&=&\mathcal{N}\Bigg(\sqrt{p_0(1-\gamma_0)^N}\frac{\hat{a}_0^{\dagger N}}{\sqrt{N!}}\nonumber\\
    &+&\sum_{j=1}^{d}\sqrt{p_j(1-\gamma_j)^N}\frac{\hat{a}_j^{\dagger N}e^{iN{\phi}_j}}{\sqrt{N!}}\Bigg)|0\rangle^{\otimes d+1},
\end{eqnarray}
with a normalization constant $\mathcal{N}$ and relative phases $\phi_j=\widetilde{\phi}_j-\widetilde{\phi}_0$. By exploiting that $|{\psi}_{N,\boldsymbol{\widetilde{\phi}}}^{(\boldsymbol{\gamma})}\rangle$ in Eq. (\ref{NOON2}) is orthogonal to support of $\hat{\sigma}_N$ as explained in Appendix A, the QFIM $F_{\rm Q}(\boldsymbol{\phi})$ is formulated as (see Appendix B for details)
\begin{align}\label{NOON_repre}
    &F_{\mathrm{Q}(j,k)}(\boldsymbol{\phi})=4\left(\sum_{j=0}^{d}p_j(1-\gamma_j)^N\right)\times\nonumber\\
    &\left\{\langle\partial_{\phi_j}\psi_{N,\boldsymbol{\widetilde{\phi}}}^{(\boldsymbol{\gamma})}|\partial_{\phi_k}\psi_{N,\boldsymbol{\widetilde{\phi}}}^{(\boldsymbol{\gamma})}\rangle-\langle\partial_{\phi_j}\psi_{N,\boldsymbol{\widetilde{\phi}}}^{(\boldsymbol{\gamma})}|\psi_{N,\boldsymbol{\widetilde{\phi}}}^{(\boldsymbol{\gamma})}\rangle\langle\psi_{N,\boldsymbol{\widetilde{\phi}}}^{(\boldsymbol{\gamma})}|\partial_{\phi_k}\psi_{N,\boldsymbol{\widetilde{\phi}}}^{(\boldsymbol{\gamma})}\rangle\right\}.
\end{align}
It can be shown that the QFIM written above is invertible, which means that every relative phases are simultaneously estimated even in the presence of photon loss. The inverse of the QFIM is analytically derived by using the Sherman-Morrison formula \cite{sherman-morrison}, and the QCRB is evaluated as
\begin{align}\label{qcrb_main}
    &\mathrm{Tr}\left\{{F}_{\rm Q}^{-1}(\boldsymbol{\phi})\right\}\nonumber\\
    &=\frac{1}{4N^2}\left\{\frac{d}{p_0(1-\gamma_0)^N}+\sum_{j=1}^{d}\frac{1}{p_j(1-\gamma_j)^N}\right\}.
\end{align}
The QCRB written above is a convex function of $\boldsymbol{p}=(p_0,\cdots,p_d)$. Thus, from the fact that a point vanishing gradient of the QCRB in Eq. (\ref{qcrb_main}) is a global optimum \cite{s.boyd}, one can analytically find the minimum QCRB over all the weighted multi-mode NOON states taking the form of Eq.~(\ref{NOON}) as 
\begin{align}\label{min_qcrb}
    &\min_{\boldsymbol{p}}\mathrm{Tr}\left[{F}_{\rm Q}^{-1}(\boldsymbol{\phi})\right]\nonumber\\
    &=\frac{1}{4N^2}\left\{\frac{\sqrt{d}}{(1-\gamma_0)^{N/2}}+\sum_{j=1}^{d}\frac{1}{(1-\gamma_j)^{N/2}}\right\}^2,
\end{align}
and the optimal weights as
\begin{align}\label{opt_p}
    p_0&=\frac{\sqrt{d}\lambda_0^N}{\sqrt{d}\lambda_0^N+\lambda_1^N+\cdots+\lambda_{d}^N},\nonumber\\
    p_j&=\frac{\lambda_j^N}{\sqrt{d}\lambda_0^N+\lambda_1^N+\cdots+\lambda_{d}^N}, \ \ \mathrm{for} \ \ j\not=0
\end{align}
with coefficients $\lambda_j=\prod_{l\not=j}(1-\gamma_l)^{1/2}$ (see Appendix B for the details). The minimum QCRB written in Eq. (\ref{min_qcrb}) successfully encapsulates the previous results in Refs.~\cite{s.hong,s.hong2,j.urrehman} as lossless case. The difference between $p_0$ and $p_j$ ($j\not=0$) arises because the reference phase is assumed to be known beforehand but the other phases are unknown. We prove that the QCRB of the multi-mode NOON states is attainable (see Appendix C). Thus, the QCRB of the multi-mode NOON states is understood as the minimum estimation uncertainty over all possible measurements.

As a classical benchmark, we consider the scheme that employs a ($d+1$)-mode coherent state described by $\otimes_{j=0}^{d}|\alpha_j\rangle$ with $\alpha_j=\sqrt{q_jN}$ ($q_0+\cdots+q_j=1$, $q_j\in[0,1]$) under the same environment \cite{r.j.glauber}, leading to the output state $\otimes_{j=0}^{d}|e^{i\widetilde{\phi}_j}\alpha_j\sqrt{1-\gamma_j}\rangle$ followed by measurement and estimation. We define the SQL as the minimum QCRB over all multi-mode pure coherent states optimized over $q_j$ and (see Appendix D for the details)
\begin{align}\label{min_qcrb_coh}
    \min_{\boldsymbol{q}}\mathrm{Tr}\left[{F}_{\rm Q,coh}^{-1}(\boldsymbol{\phi})\right]=\frac{1}{4N}\left\{\sqrt{\frac{d}{1-\gamma_0}}+\sum_{j=1}^{d}\frac{1}{\sqrt{1-\gamma_j}}\right\}^2,
\end{align}
with the optimal weights 
\begin{align}\label{opt_q}
    q_0&=\frac{\sqrt{d}\lambda_0}{\sqrt{d}\lambda_0+\lambda_1+\cdots+\lambda_{d}},\nonumber\\
    q_j&=\frac{\lambda_j}{\sqrt{d}\lambda_0+\lambda_1+\cdots+\lambda_{d}}, \ \ \mathrm{for} \ \ j\not=0.
\end{align}
When the photon loss rates $\gamma_j$ $\forall j\not=0$ are all equal to $\gamma$ except the reference mode $\gamma_{\rm ref}=\gamma_0$, the QCRB of Eq.~(\ref{min_qcrb}) is rewritten as 
\begin{align}\label{qcrb_NOON_simple}
    &\min_{\boldsymbol{p}}\mathrm{Tr}\left[{F}_{\rm Q}^{-1}(\boldsymbol{\phi})\right]\nonumber\\
    &=\frac{1}{4N^2}\left\{\frac{\sqrt{d}}{(1-\gamma_{\rm ref})^{N/2}}+\frac{d}{(1-\gamma)^{N/2}}\right\}^2,
\end{align}
which reduces to
\begin{align}
    \min_{\boldsymbol{p}}\mathrm{Tr}\left[{F}_{\rm Q}^{-1}(\boldsymbol{\phi})\right]=\frac{1}{4N^2}\frac{(\sqrt{d}+d)^2}{(1-\gamma)^{N}}
\end{align}
when $\gamma=\gamma_{\rm ref}$. For the latter case, the optimal weights are given as
\begin{align}\label{opt_p_eq}
    p_0=\frac{\sqrt{d}}{\sqrt{d}+d}, \ \ p_j=\frac{1}{\sqrt{d}+d} \ \ \mathrm{for} \ \ j\not=0,
\end{align}
and interestingly these recover the particular multi-mode NOON state proposed by Humphreys' et al. in Ref.~\cite{p.c.humphreys}. This implies that Humphreys' NOON states are optimal only when $\gamma=\gamma_{\rm ref}$, including the lossless case. On the other hand, the SQL of Eq.~(\ref{min_qcrb_coh}) is rewritten as
\begin{align}\label{qcrb_coh_simple}
    \min_{\boldsymbol{q}}\mathrm{Tr}\left[{F}_{\rm Q,coh}^{-1}(\boldsymbol{\phi})\right]=\frac{1}{4N}\left\{\sqrt{\frac{d}{1-\gamma_{\rm ref}}}+\frac{d}{\sqrt{1-\gamma}}\right\}^2,
\end{align}
when $\gamma=\gamma_j$ $\forall j\in\{1,\cdots,d\}$, and
\begin{align}
    \min_{\boldsymbol{q}}\mathrm{Tr}\left[{F}_{\rm Q,coh}^{-1}(\boldsymbol{\phi})\right]=\frac{1}{4N}\frac{(\sqrt{d}+d)^2}{1-\gamma},
\end{align}
when all $\gamma=\gamma_{\rm ref}$ for which the optimal weights $\boldsymbol{q}$ take the same form as Eq.~(\ref{opt_p_eq}). A more detailed discussion about these findings is provided below.

\subsection{Quantum advantage from two-photon NOON states}
Let us here investigate a quantum advantage of the proposed scheme using the multi-mode NOON states in comparison with the SQL under the same lossy environment. The environment that can be characterized in terms of the reference mode $\gamma_{\rm ref}$ and the other modes $\gamma=\gamma_j$ $\forall j\in[1,\cdots,d]$, which is inspired by the distinct roles that appear through Eq.~(\ref{min_qcrb}) to Eq.~(\ref{min_qcrb_coh}). To quantify a quantum advantage, we define the ratio of the QCRB for the considered scheme under study to that for the optimized multi-mode pure coherent states, written as
\begin{equation}\label{r_q}
    r_{\rm QA}=1-\frac{\mathrm{Tr}\left[{F}_{\rm Q}^{-1}(\boldsymbol{\phi})\right]}{\mathrm{Tr}\left[{F}_{\rm Q,coh}^{-1}(\boldsymbol{\phi})\right]}.
\end{equation}
A positive value of $r_{\rm QA}$ means that a methological performance of the considered scheme surpasses the SQL. In a specific case that $\gamma=\gamma_{\rm ref}$, $r_{\rm QA}$ can reduce to
\begin{equation}\label{r_qa_spec}
    r_{\rm QA}=1-\frac{1}{N}(1-\gamma)^{1-N}.
\end{equation}
Here, note that a quantum advantage is independent of the mode number $d$ and diminishes as $N$ becomes larger. On the other hand, the role of the mode number $d$ is significant when $\gamma\not=\gamma_{\rm ref}$. It can be easily shown by using Eqs.~(\ref{qcrb_NOON_simple}) and (\ref{qcrb_coh_simple}) that, in the case of $\gamma<\gamma_{\rm ref}$, the quantum advantage quantified by Eq.~(\ref{r_q}) is improved by increasing the mode number $d$. Moreover, when a multi-mode NOON state has sufficiently large $d$, the quantum advantage in Eq.~(\ref{r_q}) converges to $\lim_{d\rightarrow\infty}r_{\rm QA}=1-\frac{1}{N}(1-\gamma)^{1-N}$ as written in Eq.~(\ref{r_qa_spec}), which is independent of the photon loss rate of the reference mode. It is interesting that, for $\gamma<\gamma_{\rm ref}$, the quantum advantage of the proposed scheme is irrelevant to the photon loss rate in a reference mode.

\begin{figure}[t]
\rightline{\includegraphics[width=8.5cm]{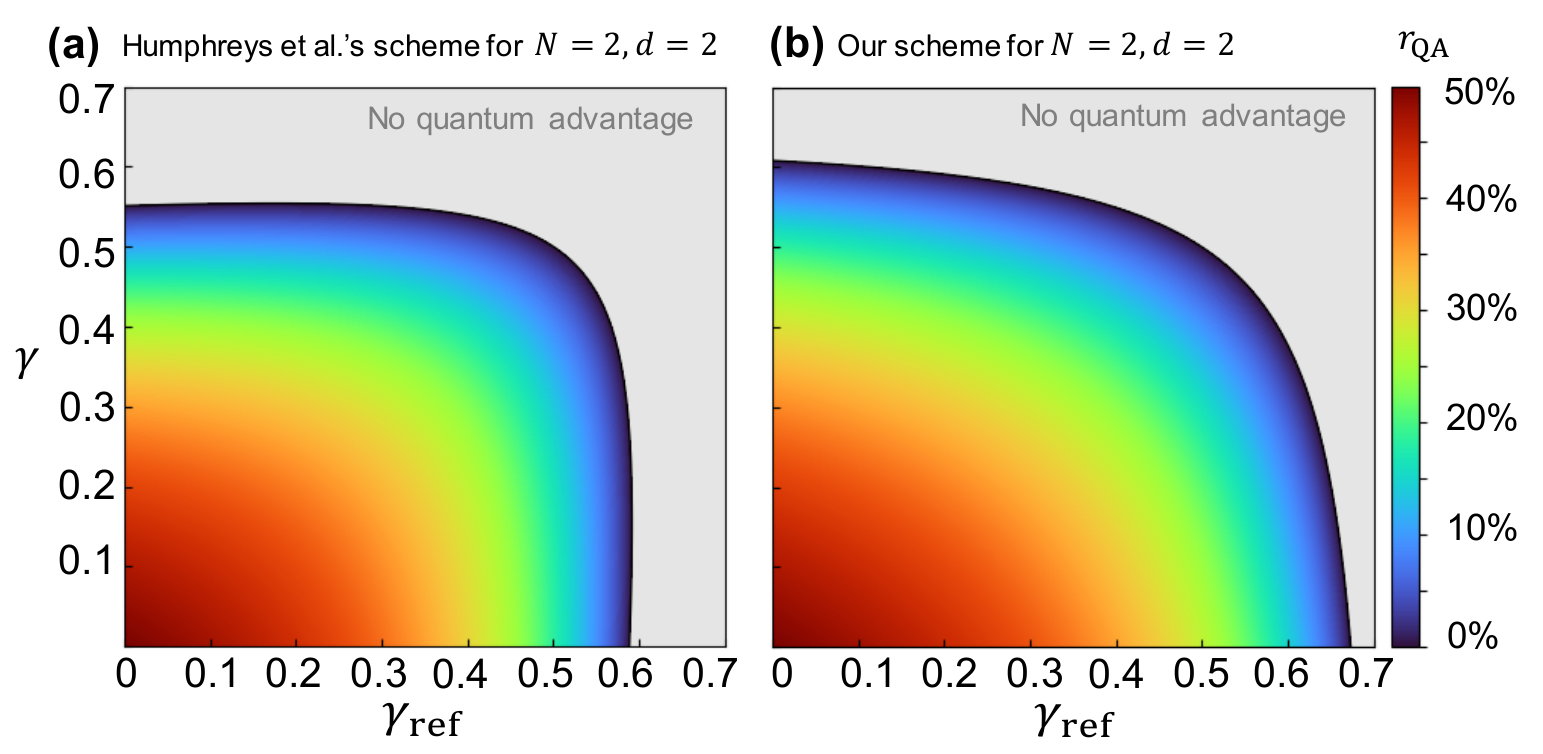}}
\caption{Quantum advantage ($r_{\rm QA}$) in terms of the photon loss rate $(\gamma_{\rm ref})$ in the reference mode and that ($\gamma$) in the other modes for (a) the scheme proposed in Ref.~\cite{p.c.humphreys} and (b) our scheme in comparison with the SQL when $N=2$ and $d=2$.}
\centering
\label{fig2}
\end{figure}

Figure~\ref{fig2} illustrates the regions of photon loss rates where a quantum advantage is achieved for the scheme proposed in Ref.~\cite{p.c.humphreys} (see Fig.~\ref{fig2}(a)) and for the scheme we propose with optimal multi-mode NOON states (see Fig.~\ref{fig2}(b)) when $N=d=2$. It is evident that the colored area of Fig.~\ref{fig2}(b) is larger than that of Fig.~\ref{fig2}(a), which implies that a multiple-phase estimation scheme proposed in this work is more robust to the photon loss than the scheme proposed in Ref.~\cite{p.c.humphreys}. This is simply because our schemes employ the weighted NOON states whose weights are optimized for individual loss rates. Also note that the proposed scheme is slightly more robust against the photon loss in the reference mode than that of the other modes. 

\begin{figure}[t]
\rightline{\includegraphics[width=8.5cm]{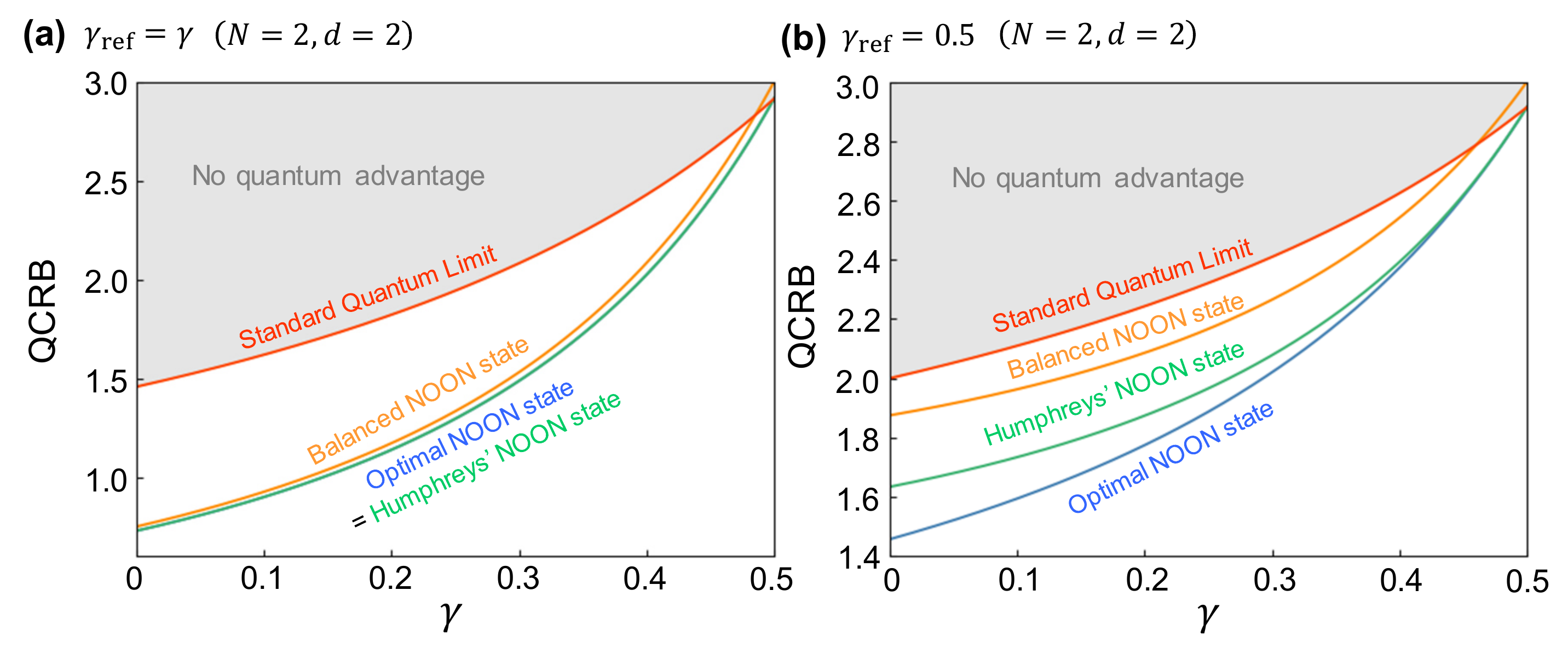}}
\caption{QCRB in terms of the photon loss rate $\gamma$ when (a) $\gamma_{\rm ref}=\gamma$ and (b) $\gamma_{\rm ref}=0.5$ for the schemes using the optimal NOON states (blue), the NOON states proposed in Ref.~\cite{p.c.humphreys} (green),  the balanced NOON states (orange), and the coherent states (red) setting the SQL. Here $N=2$ and $d=2$ are considered for all the schemes.}
\centering
\label{fig3}
\end{figure}

To look into more detailed behaviors, we take two specific values of $\gamma_{\rm ref}$ from Fig.~\ref{fig2}(b), and compare different cases including the case using balanced NOON states (i.e., $p_j=\frac{1}{d}$ $\forall j$). In Fig.~\ref{fig3}(a), we show that when $\gamma=\gamma_{\rm ref}$, our scheme is equivalent to the scheme proposed in Ref.~\cite{p.c.humphreys}, already pointed out in Eq.~(\ref{opt_p_eq}), and the balanced NOON states are also nearly optimal due to equal losses. It can be shown that the equivalence holds for any $N$ and $d$, but the near-optimality of the balanced NOON states is degraded with increasing $\gamma$. Interestingly, on the other hand, our scheme becomes significantly advantageous when $\gamma\not=\gamma_{\rm ref}$, which is clearly shown in Fig.~\ref{fig3}(b) for the case of $\gamma_{\rm ref}=0.5$ that precision difference between these two schemes is well-visible. It reveals that our scheme is better than the scheme in Ref.~\cite{p.c.humphreys} when the loss rates are unequal to each other and the balanced NOON states are far from being optimal due to unequal losses. Also note that our scheme can outperform the SQL in the lossy environment, which can be observed in Fig.~3.

\subsection{Quantum advantage from $d$-mode $N$-photon NOON states}

\begin{figure}[t]
\rightline{\includegraphics[width=8.5cm]{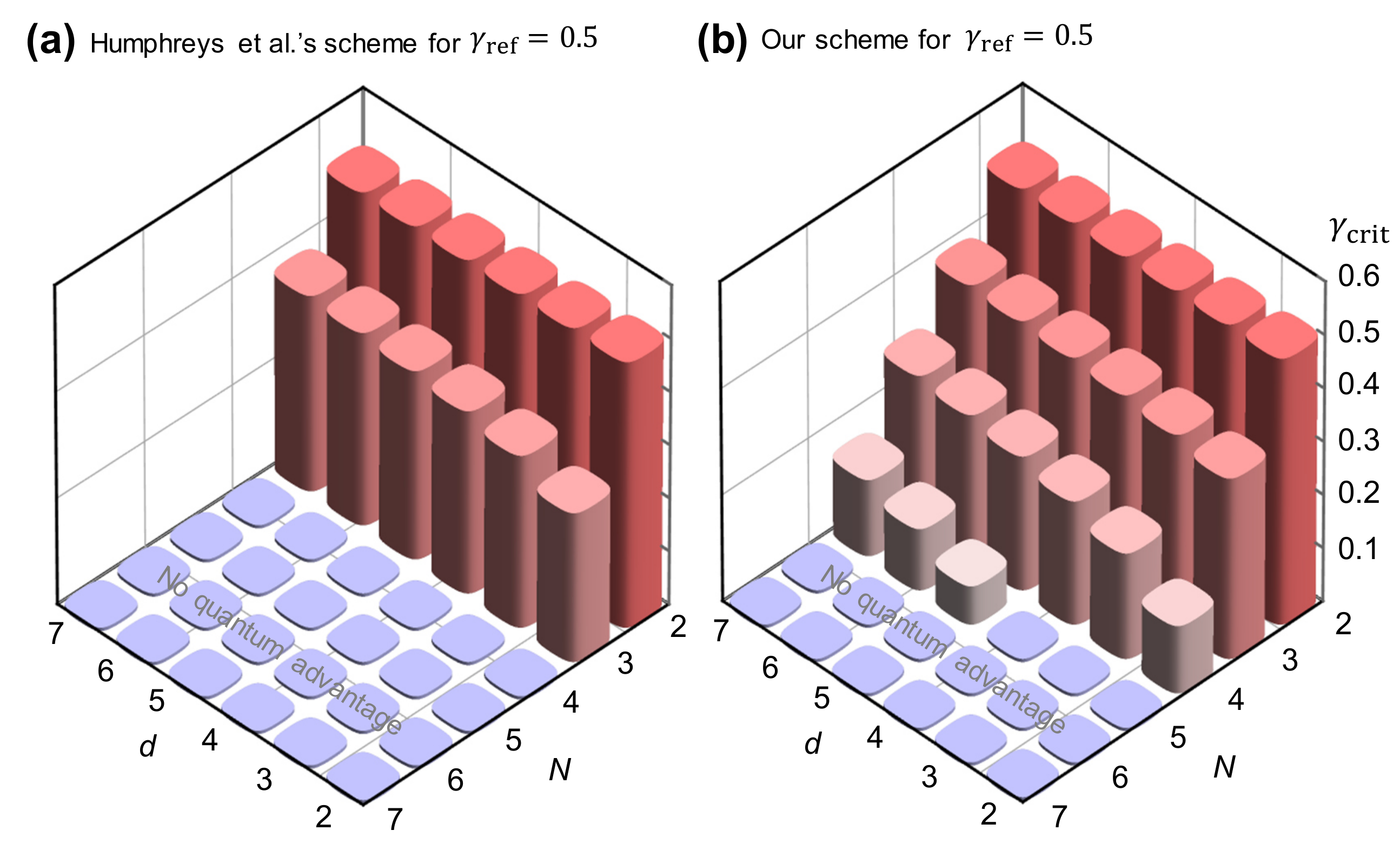}}
\caption{Critical photon loss rate $\gamma_{\rm crit}$ such that a quantum advantage remains until $\gamma\le\gamma_{\rm crit}$ exhibits the robustness of the considered scheme for $N,d\ge 2$. Blue area shows the region that $r_{\rm QA}$ is lower than zero, wherein a multi-mode NOON state does not exhibit a quantum advantage.}
\centering
\label{fig4}
\end{figure}

Here, we examine if a quantum advantage is still achievable by $d$-mode $N$-photon NOON states with $N,d>2$ and how large it is. For the purpose, we define the critical photon loss rate $\gamma_{\rm crit}$ such that a quantum advantage is observed when $\gamma_j\equiv\gamma\le\gamma_{\rm crit}$ for $j\in\{1,\cdots,d\}$ where $\gamma_{\rm ref}=0.5$. That is, $\gamma_{\rm crit}$ indicates a robustness of the scheme under study. Figures \ref{fig4}(a) and (b) present $\gamma_{\rm crit}$ for the scheme proposed in Ref.~\cite{p.c.humphreys} and $\gamma_{\rm crit}$ for our scheme, respectively, when $\gamma<\gamma_{\rm ref}=0.5$ as an example that the photon loss of the reference mode is larger than that of the other modes. It is clearly shown that the scheme proposed in this work is more robust against photon loss than Humphreys et al.'s scheme proposed in Ref.~\cite{p.c.humphreys}. Interestingly, the proposed scheme becomes more robust as the number of phases or modes increase. This means that our scheme is useful when the number of phases to be estimated is many, and offers robust multiple-phase estimation schemes even when $N<d$. Particularly for cases where $N=2$ and $d=10$, we also study general loss scenarios with arbitrary loss rates of $\gamma_j$'s. This involves simulating $10^5$ instances, where each instance's loss rates are arbitrarily generated from a uniform distribution within a specific range $[\gamma_{\rm min}, \gamma_{\rm max}]$. We find that our schemes generally perform well because our NOON states are optimized for individual loss scenarios (see Appendix F for the details).

\section{Quantum advantage in terms of the CRB}
Until the previous section, we have demonstrated that the QCRB of an optimal multi-mode NOON state can surpass the SQL. Exploiting the argument written in Appendix A together with Ref. \cite{l.pezze}, it is easy to show that a measurement achieving the QCRB has an entanglement-based projection $|{\psi}_{N,\boldsymbol{\phi}}^{(\boldsymbol{\gamma})}\rangle\langle{\psi}_{N,\boldsymbol{\phi}}^{(\boldsymbol{\gamma})}|$. However, it is known that such the measurement is experimentally challenging to implement in general \cite{s.hong2,j.urrehman}. Thus, for experimental feasibility, we need to verify whether an estimation scheme using a measurement with a linear optical structure can provides the SQL.

\begin{figure}[t]
\rightline{\includegraphics[width=8.5cm]{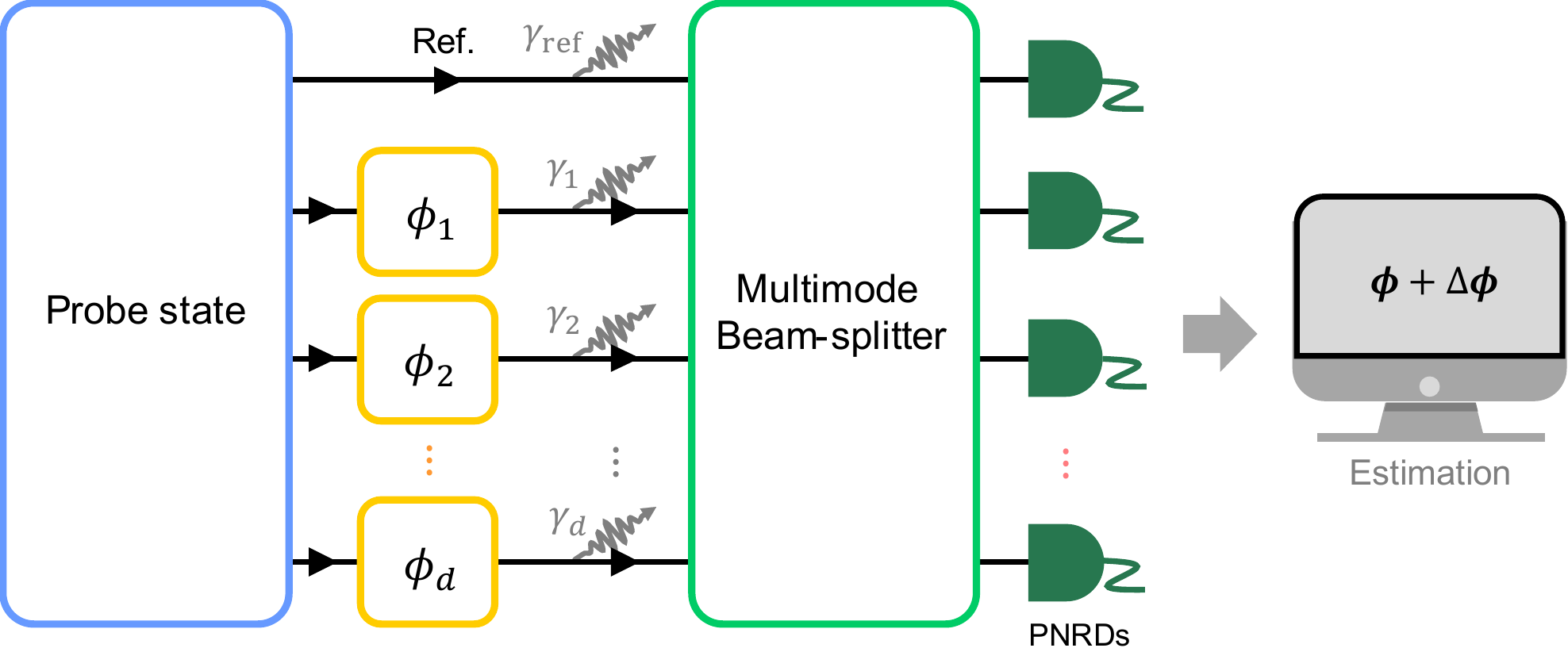}}
\caption{Multiple-phase estimation using a $d+1$-mode probe state for estimating $d$ unknown phases $\{\phi_1,\cdots,\phi_d\}$ with a measurement consisting of a multi-mode beam-splitter and PNRDs.}
\centering
\label{fig5}
\end{figure}

We consider a measurement scheme illustrated as Fig.~\ref{fig5} that employs PNRDs with a multi-mode beam-splitter being optimized in the multiple-phase estimation scheme using multi-mode NOON states, and investigate whether the associated CRB evaluated in Appendix E can surpass the SQL and how close it is to be the QCRB of Eq.~(\ref{min_qcrb}). For the optimization over all multi-mode beam-splitters, we exploit Clement's configuration \cite{w.r.clements} to represent a general multi-mode beam-splitter as a $(d+1)$-mode transformation 
\begin{equation}\label{clements_con}
    \mathrm{U}=\mathrm{D}\prod_{j}\mathrm{V}_j(\theta_j,\chi_j).
\end{equation}
Here, $\mathrm{D}$ is a diagonal unitary matrix and $\mathrm{V}_j(\theta_j,\chi_j)$ is a block unitary matrix composed of a $(d-1)$-dimensional identity matrix and 2 dimensional unitary matrix with real parameters $\theta_j$ and $\chi_j$, which is implemented by a beam-splitter and a phase shifter. Note that $\mathrm{D}$ consists of redundant phases disappeared from the CRB. In other words, the optimization over all possible multi-mode beam-splitters is equivalent to that over all the real parameters in Eq. (\ref{clements_con}).

\begin{figure}[t]
\rightline{\includegraphics[width=8.5cm]{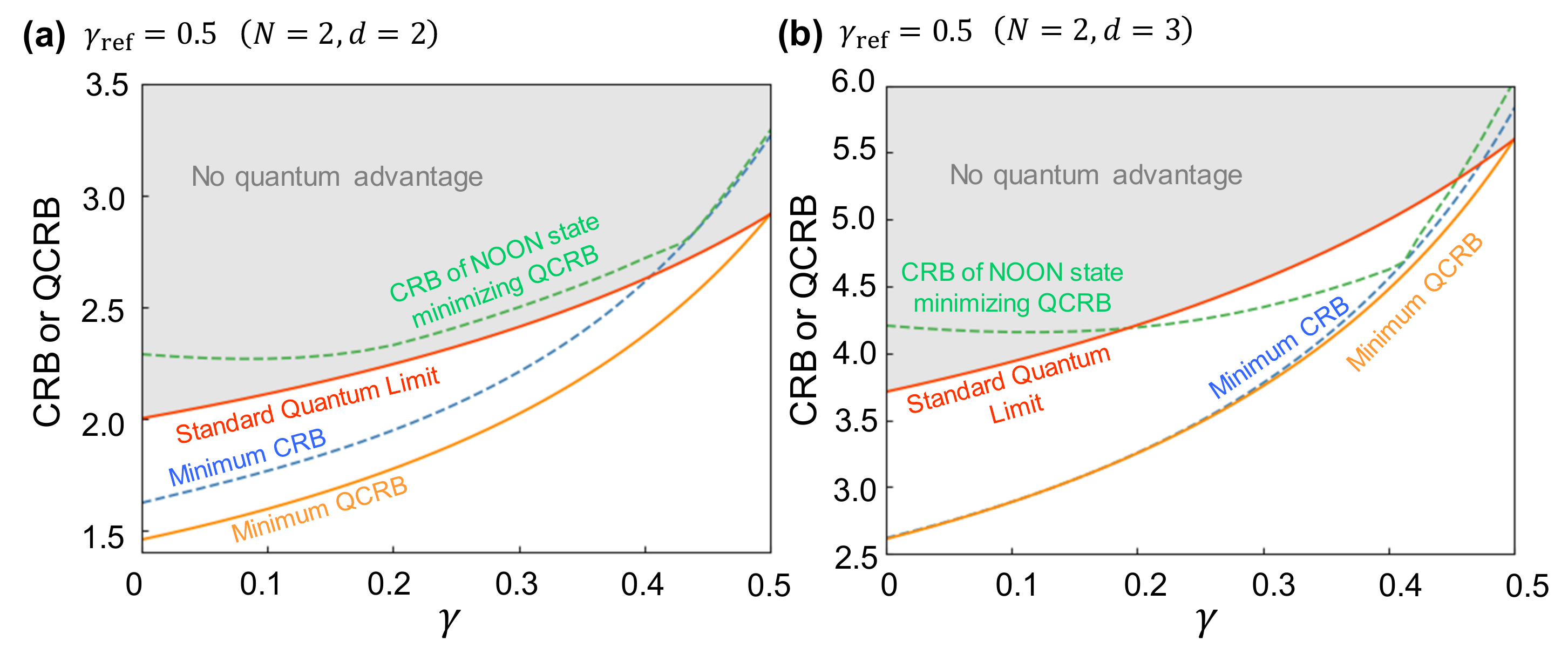}}
\caption{CRB of our schemes with PNRDs performed through the multi-mode beam-splitter are compared with the QCRB of Eq.~(\ref{expl_NOON}) (orange) and the SQL of Eq.~(\ref{min_qcrb}) (red), in terms of $\gamma$ when $\gamma_{\rm ref}=0.5$ for (a) $N=2$, $d=2$ and (b) $N=2$, $d=3$. The PNRDs are applied to not only the multi-mode NOON states having been optimized with respect to Eq.~(\ref{expl_NOON}) (green), but also the case that both the multi-mode NOON state and multi-mode beam-splitter are simultaneously optimized to maximize the CRB (blue).}
\centering
\label{fig6}
\end{figure}

Now, we use the configuration introduced in Eq. (\ref{clements_con}) for the optimal multi-mode beam-splitter and evaluate the two cases: (i) we optimize the multi-mode beam-splitter for the multi-mode NOON states having minimized the QCRB of Eq.~(\ref{min_qcrb}), and (ii) we optimize simultaneously the multi-mode beam-splitter and the NOON state to minimize the CRB of Eq.~(\ref{fim_def}). The results are shown in Fig.~\ref{fig6} for $N=2$ and $d=2,3$ when $\gamma_{\rm ref}=0.5$. It is clear that the considered measurement scheme provides mostly inferior performance when applied to the NOON states that have been optimized with respect to the QCRB. However, when both the multi-mode beam-splitter and the NOON states are simultaneously optimized to minimize the CRB, their precision becomes significantly improved and even nearly optimal with respect to the minimum QCRB. Experimental structure of optimal multi-mode beam-splitter depends on the number of modes $d$. The minimum CRB of Fig.~\ref{fig6}(a) is attained by a multi-mode beam-splitter with unbalanced ratio, but that of Fig.~\ref{fig6}(b) is exhibited from a balanced one. \\

\section{Conclusion}
We have proposed and studied an optimal scheme of multiple-phase estimation using the weighted multi-mode NOON states under a lossy environment. We have shown that the proposed scheme surpasses the SQL even in the presence of loss, and is more robust to loss than the scheme proposed in Ref.~\cite{p.c.humphreys} that employs only a particular type of multi-mode NOON state. These behaviors persist when the numbers of modes and photons increase, and especially a quantum advantage even increases with the number of modes when the reference mode is lossier than the signal modes carrying the multi-mode NOON states. When the proposed scheme is equipped with a measurement setting configurable by using a multi-mode beam-splitter and PNRDs, a quantum advantage is still achieved although sub-optimal, which provides a practical means in multiple-phase estimation under a lossy environment. The sub-optimality may be due to that eigenvectors composing an SLD operator is different from bases of the measurement with optimized multi-mode beam-splitter, which will be further discussed in the future work. We believe the results obtained and the methodology used in this work will motivate various relevant studies in robust quantum metrology under lossy and noisy environments. 

An interesting future study would be to apply our methodology to mitigate other types of noises, e.g., the phase-diffusion \cite{m.t.dimario}, or to estimate other types of multiple parameters of interest, e.g., absorption parameters \cite{a.karsa}. One may also investigate if similar advantage and robustness can be obtained by employing other strategies such as reference configurations \cite{a.z.goldberg} or post-selection \cite{drm}. Another application of this work can be made to estimate a linear function of multiple parameters such as distributed quantum sensing~\cite{s.-r.zhao,x.guo,j.liu3,d.-h.kim}.

\section*{Appendix A. Phase-encoded multi-mode NOON states under photon loss}
Here we derive the explicit form of a phase-encoded multi-mode NOON state in Eq.~(\ref{expl_NOON}) exposed to photon loss. We first note that a quantum channel $\Lambda_{\gamma_j}(\cdot)$ is described by
\begin{align}\label{lambda}
    \Lambda_{\gamma_j}(\cdot)=\mathrm{Tr}_{e_j}\left\{\hat{B}_j(\cdot)_{a_j}\otimes |0\rangle\langle 0|_{e_j}\hat{B}_j^\dagger\right\}
\end{align}
with the vacuum state $|0\rangle$ in the environment mode $j$ and unitary operator $\hat{B}_j$ with respect to a mode transformation $\hat{B}_j\hat{a}_j^\dagger \hat{B}_j^\dagger=\sqrt{1-\gamma_j}\hat{a}_j^\dagger+\sqrt{\gamma_j}\hat{e}_j^\dagger$ \cite{s.m.barnett}, where $\mathrm{Tr}_{e_j}\left\{\cdot\right\}$ is the partial trace over the $j$-th environment system. 

\begin{widetext}
\noindent From Eq.~(\ref{lambda}), It is straightforward to derive three terms $\Lambda_{\gamma_j}(|N\rangle\langle N|)$, $\Lambda_{\gamma_j}(|N\rangle\langle 0|)$, and $\Lambda_{\gamma_j}(|0\rangle\langle 0|)$ as
\begin{align}
    \Lambda_{\gamma_j}(|N\rangle\langle N|)&=\sum_{k=0}^{N}\frac{N!}{(N-k)!k!}(1-\gamma_j)^{N-k}\gamma_j^k|N-k\rangle\langle N-k|,\nonumber\\
    \Lambda_{\gamma_j}(|N\rangle\langle 0|)&=(1-\gamma_j)^{N/2}|N\rangle\langle 0|, \ \ \Lambda_{\gamma_j}(|0\rangle\langle 0|)=|0\rangle\langle 0|.
\end{align}
Substituting the above terms in Eq.~(\ref{lossy_NOON}), the output state is derived as
    \begin{align}\label{det_vers}
        \hat{\rho}_{N,\boldsymbol{\phi}}&=\left\{\sum_{j=0}^{d}p_j(1-\gamma_j)^N\right\}|{\psi}_{N,\boldsymbol{\phi}}^{(\boldsymbol{\gamma})}\rangle\langle {\psi}_{N,\boldsymbol{\phi}}^{(\boldsymbol{\gamma})}|\nonumber\\
        &+\underbrace{\sum_{j=0}^{d}p_j\left\{\sum_{r=1}^{N}\frac{N!}{r!(N-r)!}(1-\gamma_j)^{N-r}\gamma_j^r\frac{\hat{a}_j^{\dagger N-r}|0\rangle\langle 0|^{\otimes d}\hat{a}_j^{N-r}}{(N-r)!}\right\}}_{\hat{\sigma}_N}.
    \end{align}

\section*{Appendix B. Evaluation of the minimum QCRB}
We begin the detailed evaluation by introducing methodology to evaluate the QFIM, which was proposed in Refs.~\cite{j.liu2,m.g.a.paris,j.liu_sld}. Let us consider a density operator $\hat{\rho}_{\boldsymbol{\phi}}=\sum_{j}\lambda_j|\lambda_j\rangle\langle\lambda_j|$ with eigenvalues $\lambda_j$ and corresponding eigenvectors $|\lambda_j\rangle$. If $\hat{\rho}_{\boldsymbol{\phi}}$ is full rank, then an SLD operator defined in Eq. (\ref{qfim}) is decomposed as \cite{m.g.a.paris}
\begin{eqnarray}
    \hat{L}_{a}=\sum_{j}\frac{\partial_{\phi_a}\lambda_j}{\lambda_j}|\lambda_j\rangle\langle\lambda_j|+2\sum_{j\not=k}\frac{\lambda_j-\lambda_k}{\lambda_j+\lambda_k}\langle\lambda_k|\partial_{\phi_a}\lambda_j\rangle|\lambda_k\rangle\langle\lambda_j|.
\end{eqnarray}
The above decomposition admits Eq. (\ref{qfim}) to be written as \cite{j.liu2,j.liu_sld}
\begin{eqnarray}\label{qfim_form}
    F_{\mathrm{Q}(a,b)}(\boldsymbol{\phi})&=&\sum_{j}\frac{(\partial_{\phi_a}\lambda_j)(\partial_{\phi_b}\lambda_j)}{\lambda_j}+4\sum_{j}\lambda_j\mathrm{Re}\left\{\langle\partial_{\phi_a}\lambda_j|\partial_{\phi_b}\lambda_j\rangle\right\}\nonumber\\&-&8\sum_{j,k}\frac{\lambda_j\lambda_k}{\lambda_j+\lambda_k}\mathrm{Re}\left\{\langle\partial_{\phi_a}\lambda_j|\lambda_k\rangle\langle\lambda_k|\partial_{\phi_b}\lambda_j\rangle\right\}.
\end{eqnarray}

Based on the QFIM formalism introduced above, we analytically evaluate the minimum QCRB of $d+1$-mode NOON states. The non-negative operator $\hat{\sigma}_N$ in Eq.~(\ref{det_vers}) is diagonalized by vectors of the Fock basis whose photon number is strictly less than $N$. Meanwhile, $|{\psi}_{N,\boldsymbol{\phi}}^{(\boldsymbol{\gamma})}\rangle$ in Eq.~(\ref{det_vers}) is a linear combination of the Fock basis vectors with photon number equal to $N$. Thus, the support of $\hat{\sigma}_N$ in Eq.~(\ref{det_vers}) is orthogonal to $|{\psi}_{N,\boldsymbol{\phi}}^{(\boldsymbol{\gamma})}\rangle$, and $\hat{\rho}_{N,\boldsymbol{\phi}}$ is in the spectral decomposition with the eigenvectors $|0\rangle,\cdots,|N-1\rangle$ together with $|{\psi}_{N,\boldsymbol{\phi}}^{(\boldsymbol{\gamma})}\rangle$. We can obtain Eq.~(\ref{NOON_repre}) from Eq. (\ref{qfim_form}), which is employed to derive components of the QFIM
\begin{align}
    {F}_{\mathrm{Q}(a,b)}(\boldsymbol{\phi})=4N^2\left\{\sqrt{p_ap_b(1-\gamma_a)^{N}(1-\gamma_b)^{N}}\delta_{ab}-\frac{p_ap_b(1-\gamma_a)^N(1-\gamma_b)^N}{p_0(1-\gamma_0)^N+\cdots+p_{d}(1-\gamma_d)^{N}}\right\}.
\end{align}
By using the Sherman-Morrison formula \cite{sherman-morrison}, the QCRB is evaluated as 
\begin{align}\label{qcrb_der}
    \mathrm{Tr}\left\{{F}_{\rm Q}^{-1}(\boldsymbol{\phi})\right\}=\frac{1}{4N^2}\left\{\frac{d}{p_0(1-\gamma_0)^N}+\sum_{j=1}^{d}\frac{1}{p_j(1-\gamma_j)^N}\right\},
\end{align}
which is the convex function of $\boldsymbol{p}=(p_0,\cdots,p_{d})$. Since a feasible set of $\boldsymbol{p}$ is convex and $\mathrm{Tr}\left\{F_{\rm Q}^{-1}(\boldsymbol{\phi})\right\}$ is a convex function, a global minimum is represented as a point where gradient of the QCRB in Eq.~(\ref{qcrb_der}) is vanished \cite{s.boyd}. Also, normalization condition $p_0+p_1+\cdots+p_d=1$ leads us to 
\begin{eqnarray}
    \frac{\partial}{\partial p_j}\mathrm{Tr}\left\{F_{\rm Q}(\boldsymbol{\phi})\right\}=\frac{1}{4N^2}\left\{\frac{d}{p_0^2(1-\gamma_0)^N}-\frac{1}{p_j^2(1-\gamma_j)^N}\right\}
\end{eqnarray}
with vanishing-gradient conditions
\begin{eqnarray}
    p_j=\frac{1}{\sqrt{d}}\frac{(1-\gamma_0)^{N/2}}{(1-\gamma_j)^{N/2}}p_0, \ \ \forall j\in\{1,\cdots,d\}.
\end{eqnarray}
Hence, we obtain optimal weights in Eq.~(\ref{opt_p}) from $p_j$ written above together with the normalization condition. Consequently, the minimum QCRB is evaluated as Eq.~(\ref{min_qcrb_coh}) by the optimal weights.

\section*{Appendix C. Attainability of the QCRB}
It was shown that the necessary and sufficient condition that the QCRB of a quantum state $\hat{\rho}$ is attained by a measurement is \cite{j.liu2}
\begin{align}\label{nes_suf_cond}
\mathrm{Tr}\left(\hat{\rho}[\hat{L}_a,\hat{L}_b]\right)=0,
\end{align}
where $\hat{L}_a$ and $\hat{L}_b$ are symmetric logarithm derivative operators. By using the above proposition, we show that there is a measurement attaining the QCRB in Eq.~(\ref{qcrb_der}). From the output state described by Eq.~(\ref{det_vers}), the trace of the left hand side in Eq.~(\ref{nes_suf_cond}) is written as linear combination of $\langle{\psi}_{N,\boldsymbol{\phi}}^{(\boldsymbol{\gamma})}|[\hat{L}_a,\hat{L}_b]|{\psi}_{N,\boldsymbol{\phi}}^{(\boldsymbol{\gamma})}\rangle$ and $\langle0|^{\otimes d+1}\hat{a}_j^{N-r}[\hat{L}_a,\hat{L}_b]\hat{a}_j^{\dagger N-r}|0\rangle^{\otimes d+1}$ with $r\not=0$. First, since $|{\psi}_{N,\boldsymbol{\phi}}^{(\boldsymbol{\gamma})}\rangle$ in Eq.~(\ref{det_vers}) is a multi-mode NOON state, there are symmetric logarithm derivative operators $\hat{L}_a$ and $\hat{L}_b$ such that $\langle{\psi}_{N,\boldsymbol{\phi}}^{(\boldsymbol{\gamma})}|[\hat{L}_a,\hat{L}_b]|{\psi}_{N,\boldsymbol{\phi}}^{(\boldsymbol{\gamma})}\rangle=0$ \cite{p.c.humphreys}. Second, as we consider $\hat{L}_a$ and $\hat{L}_b$ of the multi-mode NOON state $|{\psi}_{N,\boldsymbol{\phi}}^{(\boldsymbol{\gamma})}\rangle$, their supports are orthogonal to $\hat{a}_j^{\dagger N-r}|0\rangle^{\otimes d+1}$ as $r\not=0$. Thus, $\langle0|^{\otimes d+1}\hat{a}_j^{N-r}[\hat{L}_a,\hat{L}_b]\hat{a}_j^{\dagger N-r}|0\rangle^{\otimes d+1}$ also become zero. Finally, the symmetric logarithm derivative operators satisfy the equality in Eq.~(\ref{nes_suf_cond}), which leads us to understand the attainability of the QCRB.

\section*{Appendix D. Evaluation of the SQL}
We first remind that total uncertainty $(\Delta\boldsymbol{\phi})^2$ is equal to $(\Delta\boldsymbol{\phi})^2=\sum_{j=1}^{d}(\Delta(\widetilde{\phi}_j-\widetilde{\phi}_0))^2=d(\Delta\widetilde{\phi}_0)^2+\sum_{j=1}^{d}(\Delta\widetilde{\phi}_j)^2$, and each $(\Delta\widetilde{\phi}_j)^2$ with respect to a lossy coherent state $|e^{i\widetilde{\phi}_j}\sqrt{1-\gamma_j}\alpha_j\rangle$ admits QCRB inequality $(\Delta\widetilde{\phi}_j)^2\ge\frac{1}{4|\alpha_j|^2(1-\gamma_j)}$ \cite{c.oh}. Thus, lower bound of the total uncertainty is evaluated as
\begin{eqnarray}\label{sql_pure_coh}
(\Delta\boldsymbol{\phi})^2&=&d(\Delta\widetilde{\phi}_0)^2+\sum_{j=1}^{d}(\Delta\widetilde{\phi}_j)^2\nonumber\\
&\ge&\frac{d}{4|\alpha_0|^2(1-\gamma_0)}+\frac{1}{4|\alpha_1|^2(1-\gamma_1)}+\cdots+\frac{1}{4|\alpha_d|^2(1-\gamma_d)}\nonumber\\
&=&\frac{1}{4N}\left\{\frac{d}{q_0(1-\gamma_0)}+\frac{1}{q_1(1-\gamma_1)}+\cdots+\frac{1}{q_d(1-\gamma_d)}\right\},
\end{eqnarray}
where we parameterize $|\alpha_j|$ as $|\alpha_j|=\sqrt{q_jN}$ with $q_0+\cdots+q_d=1$ and $q_j\ge 0$ $\forall j\in\{0,\cdots,d\}$. It is straightforward to show that Hessian matrix of the lower bound in Eq. (\ref{sql_pure_coh}) is positive-definite for any points $(q_0,\cdots,q_d)$, which means that a point that vanishes the gradient yields the global minimum \cite{s.boyd}. Consequently, the minimum QCRB written in Eq. (\ref{min_qcrb_coh}) is obtained by substituting the vanishing-gradient point to the lower bound in Eq. (\ref{sql_pure_coh}).

\section*{Appendix E. Measurement probabilities for the CRB}
For evaluating the CRB, we first derive the FIM in Eq.~(\ref{fim_def}) by deriving the measurement probabilities $P_{\boldsymbol{l}}$. Let us start from expressing the measurement probability as
\begin{align}\label{p_l}
    P_{\boldsymbol{l}}=\left|\langle \boldsymbol{l}|\hat{U}_{\rm BS}\hat{\rho}_{N,{\boldsymbol{\phi}}}\hat{U}_{\rm BS}|\boldsymbol{l}\rangle\right|^2,
\end{align}
where $|\boldsymbol{l}\rangle=|l_0\rangle\otimes|l_1\rangle\otimes\cdots\otimes|l_d\rangle$ is a multi-mode Fock basis vector with respect to measurement outcomes $\boldsymbol{l}=(l_0,\cdots,l_d)$, and $\hat{U}_{\rm BS}$ is a unitary operator of a multi-mode beam-splitter. By representing the unitary transformation with $\hat{U}_{\rm BS}$ as a scattering matrix $\mathrm{U}$, the above measurement probability is derived as
\begin{align}\label{p_l_rewr}
    P_{\boldsymbol{l}}=\left|\sqrt{\frac{N!}{\prod_{j=0}^{d}l_j!}}\left\{\sum_{j=0}^{d}p_j^{1/2}(1-\gamma_j)^{N/2}e^{iN\phi_j}\mathrm{U}_{j0}^{*l_0}\mathrm{U}_{j1}^{*l_1}\cdots\mathrm{U}_{jd}^{*l_d}\right\}\right|^2+\langle \boldsymbol{l}|\hat{U}_{\rm BS}\hat{\sigma}_N\hat{U}_{\rm BS}|\boldsymbol{l}\rangle.
\end{align}
We note that the second term of $P_{\boldsymbol{l}}$ is independent of the phases $\boldsymbol{\phi}$. Thus, the CRB is only determined by the first term of Eq.~(\ref{p_l_rewr}) by the definition in Eq.~(\ref{fim_def}). 

\begin{figure*}[t]
\centerline{\includegraphics[width=11.5cm]{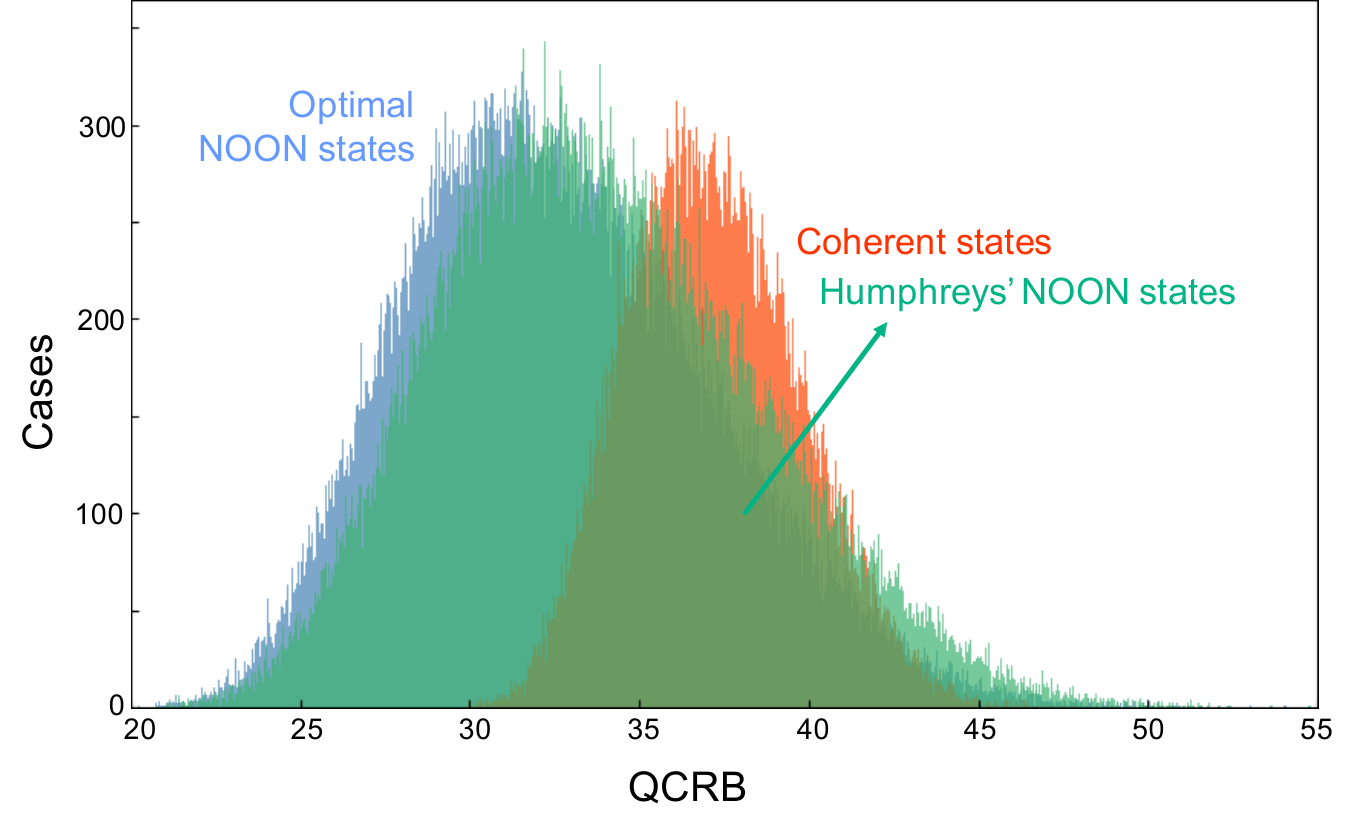}}
\caption{Histogram of multi-mode NOON states' QCRB in cases with arbitrarily chosen photon loss rates $\gamma_j\in[0.2,0.6]$.}
\centering
\label{fig7}
\end{figure*}

\section*{Appendix F. Investigation of QCRB in cases with general photon loss rates}
For considering general loss cases, we consider $10^5$ instances with arbitrarily chosen photon loss rates from a uniform distribution within a specific range $[\gamma_{\rm min}=0.2,\gamma_{\rm max}=0.6]$, particularly for cases where $N=2$ and $d=10$. For individual instances, we calculate the QCRBs for our optimal NOON states, Humphreys et al.'s NOON states \cite{p.c.humphreys}, and coherent states, and illustrate the results in Fig.~\ref{fig7}. We observe that the distribution of the optimal multi-mode NOON states is more left-skewed than that of the multi-mode NOON states from Ref. \cite{p.c.humphreys}, although both distributions have the common minimum and maximum when $\gamma_j=\gamma_{\rm min}$ and $\gamma_j=\gamma_{\rm max}$ for all $j$'s, respectively. It means that estimation schemes are more robust to photon loss in general since our NOON states are optimized individually with respect to photon loss rates. We also note that the distribution of NOON states (both ours and Humphreys' et al.'s \cite{p.c.humphreys}) is wider than that of multi-mode pure coherent states. This implies that multi-mode NOON states are more sensitive against photon loss than the coherent states.
\end{widetext}

\section*{Acknowledgement}
This work was partly supported by National Research Foundation of Korea (NRF) grant funded by the Korea government (MSIT) (2022M3K4A1097123), Institute for Information \& communications Technology Planning \& Evaluation (IITP) grant funded by the Korea government (MSIT) (2020-0-00947, RS-2023-00222863), and the KIST research program (2E32941).

\end{document}